\documentclass[11pt,showpacs,preprintnumbers,amsmath,amssymb,prd,nofootinbib,superscriptaddress]{revtex4-2}

\usepackage{dcolumn}
\usepackage{bm}
\usepackage{ifpdf}
\usepackage{hyperref}
\usepackage{bm}
\usepackage{xcolor,color,graphicx,graphics,physics}
\usepackage[spanish,english]{babel}
\usepackage[latin1]{inputenc}
\usepackage[OT1]{fontenc}
\usepackage{latexsym,amssymb,amsmath,amsfonts, slashed}
\usepackage{makeidx}
\usepackage{epsfig,subfigure}
\usepackage{natbib}
\usepackage{epstopdf}
\usepackage{mathrsfs}
\usepackage{hyperref}
\hypersetup{colorlinks=true, linkcolor=blue, citecolor=blue, urlcolor=blue}
\usepackage{enumerate}
\usepackage{cancel}

\usepackage{braket}
\usepackage{fixmath}

\everymath{\displaystyle}
\usepackage{graphicx}

\usepackage[T1]{fontenc}
\usepackage{amsmath}
\usepackage{amssymb}
\usepackage{graphicx}
\usepackage{xcolor}

\newcommand{\bea}{\begin{eqnarray}}
\newcommand{\eea}{\end{eqnarray}}

\newcommand{\til}{\text{\textasciitilde}}
\ExplSyntaxOn
\NewDocumentCommand{\RN}{m}
 {
  \textup{ \int_to_Roman:n { #1 } }
 }
\ExplSyntaxOff

\newcommand{\orcid}[1]{\href{https://orcid.org/#1}{\includegraphics[width=10pt]{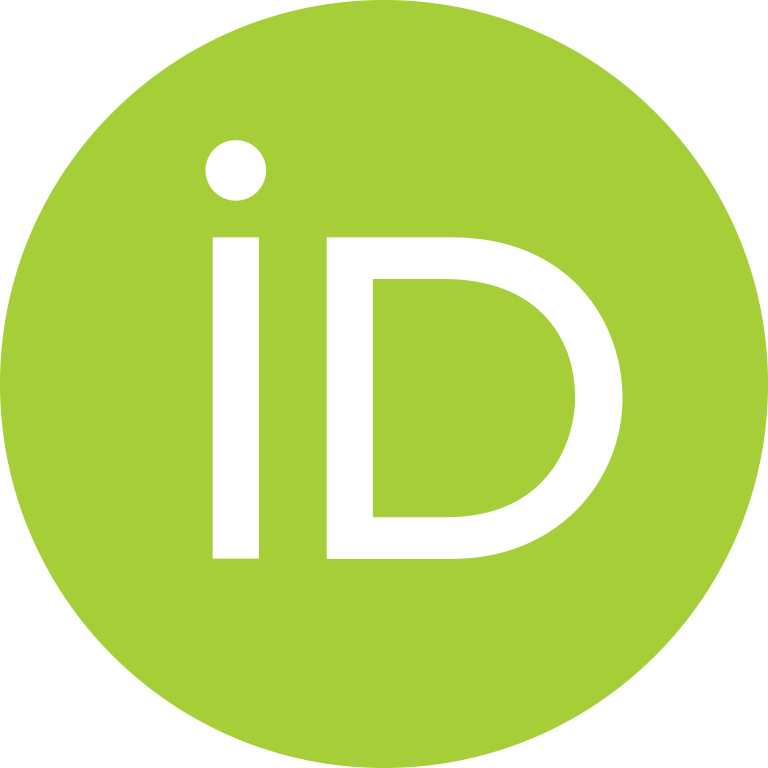}}}

\begin{document}

\title{Gravitational Compton scattering at zero and finite temperature}

\author{L. A. S. Evangelista \orcid{0009-0002-3136-2234}}
\email{lucassouza@fisica.ufmt.br }
\affiliation{Programa de P\'{o}s-Gradua\c{c}\~{a}o em F\'{\i}sica, Instituto de F\'{\i}sica,\\ 
Universidade Federal de Mato Grosso, Cuiab\'{a}, Brasil}

\author{A. F. Santos \orcid{0000-0002-2505-5273}}
\email{alesandroferreira@fisica.ufmt.br}
\affiliation{Programa de P\'{o}s-Gradua\c{c}\~{a}o em F\'{\i}sica, Instituto de F\'{\i}sica,\\ 
Universidade Federal de Mato Grosso, Cuiab\'{a}, Brasil}

\begin{abstract}

The Lagrangian formulation of Gravitoelectromagnetism (GEM) theory is considered. GEM is a gravitational theory constructed based on the similarities between gravity and electromagnetism. In this framework, we investigate gravitational Compton scattering by calculating its cross section at both zero and finite temperatures. Thermal effects are introduced via the Thermo Field Dynamics formalism. Some comparisons between GEM theory and QED have been developed. The limits of high temperature have been analyzed.

\end{abstract}

\maketitle

\section{Introduction}

The Standard Model (SM) is currently the quantum theory that describes the interactions between known fundamental particles. It is renowned for unifying the three main forces of nature (electromagnetic, strong, and weak), enabling mathematical calculations that have been used to predict particles that were purely hypothetical until then. One example is Quantum Electrodynamics (QED), which precisely describes the interaction of charged particles, such as electrons and positrons, allowing a quantum description of Maxwell's electromagnetic theory. This theory is of great value to particle physics due to its accuracy concerning experimental results. However, although the Standard Model describes the primary interactions in quantum nature, it does not unify the gravitational force with the other three fundamental forces. This is because the gravitational force is a classical theory, while the other three have a quantum nature. As a result, various studies have been developed to seek a theory of quantum gravity. One of these theories is Gravitoelectromagnetism (GEM), which aims to describe gravitons, the hypothetical fundamental particles of gravity, using concepts from electromagnetic theory.

GEM is a theory that seeks to describe gravity through a framework similar to electromagnetism. In a certain limit, it can describe the same phenomena as General Relativity (GR). Its development dates back to the period before Einstein's general relativity postulates. The first ideas were discussed by Faraday \cite{Faraday} and Maxwell \cite{Maxwell}. In 1893, Heaviside proposed ways to explain the rapid advancement of Mercury's perihelion, developing a set of equations similar to Maxwell's equations of electromagnetic theory \cite{heaviside1893gravitational}. However, this attempt was unsuccessful, as Lorentz later demonstrated that this new force, called the gravitomagnetic force, was too weak to explain such a phenomenon \cite{lorentz1899considerations}. Despite this, these concepts seemed to be related in some way. In 1918, Lense and Thirring showed that the existence of a gravitomagnetic field, generated by a massive rotating body, could affect the orientation of the orbit of another body in its vicinity, and this effect became known as the Lense-Thirring effect \cite{LT}. Starting in the 1950s, new ideas emerged based on the decomposition of the Weyl tensor into its gravitoelectric and gravitomagnetic components \cite{matte1953nouvelles, bel1958radiation}. This gravitational theory relies on two key assumptions: (i) a gravitomagnetic field is associated with moving masses, and (ii) the gravitational field propagates at the speed of light. 

Although GEM theory can be derived through various methods, this discussion will focus on the approach based on the decomposition of the Weyl tensor. Using this approach, a Lagrangian formulation for GEM theory has been developed \cite{Jair}. In this formalism, the gravitational field is described by the tensor potential $A^{\mu\nu}$. From the Lagrangian formulation, various studies involving the graviton can be explored. For example, there are studies discussing gravitational electron-positron scattering \cite{jesus2022gravitational}, gravitational Bhabha scattering \cite{santos2017gravitational}, gravitational M\"{o}ller scattering \cite{alesandrogravitacional}, gravitational Casimir effect \cite{Casimir}, among others. In this work, we will use the Lagrangian formulation of GEM to investigate gravitational Compton scattering. This description is particularly valuable because, while GEM is analogous to GR in the weak-field limit \cite{Mashhon}, allowing physical processes to be described by either theory, its Lagrangian formulation also enables the quantization of the theory. This facilitates the study of graviton interactions with fermions and photons. Although it might appear that we are sidelining GR in favor of GEM, our aim is to demonstrate that GEM is consistent with GR and can effectively describe both classical and quantum phenomena \cite{santos2017gravitational, farrugia2020gravitoelectromagnetism, chatzistavrakidis2020torsion}. For a more detailed exploration of the relationship between GEM and GR, we direct readers to Refs. \cite{alesandrogravitacional, Mashhon, bakopoulos2016gravitoelectromagnetism}. In this context, GEM proves to be a more suitable framework for describing quantum particle interactions.  It is important to note that GEM describes gravity in a way analogous to GR in the weak-field limit. However, it is not suited for describing strong gravitational phenomena, such as black holes.

Compton scattering is a fundamental process in physics where a photon collides with an electron. The scattering was first experimentally observed by Arthur Compton in 1923 \cite{Compton}, confirming the particle-like nature of photons and providing direct evidence for the quantum theory of light. Compton scattering plays a crucial role in fields such as astrophysics, quantum electrodynamics, contributing significantly to our understanding of electromagnetic interactions at the microscopic level \cite{cabral}. Here, the gravitational version of this scattering is investigated. Starting from the hypothesis that gravitons exist as the quanta of the gravitational field, their interaction with fermions emerges as a particularly compelling subject for theoretical investigation. This involves replacing the photon with a graviton in the scattering process, i.e., $g+f\xrightarrow{}g+f$. The main aim of this work is to calculate the cross section for this gravitational scattering at zero and finite temperature. Thermal effect will be introduced using the Thermo Field Dynamics (TFD) formalism.

TFD is a real-time quantum field theory at finite temperature constructed from the idea that the statistical average of an arbitrary operator is equivalent to its vacuum expectation value \cite{Umezawa1, Umezawa2, Book, Umezawa22, Khanna1, Khanna2, Santana1, Santana2}. In this formalism, a thermal vacuum state $|0(\beta)\rangle$ is constructed. However, two key ingredients are required: doubling the Hilbert space and implementing the Bogoliubov transformation. Although there are other methods for introducing temperature into systems, such as the imaginary-time Matsubara formalism \cite{matsubara} and the real-time Closed-Time Path (CTP) approach \cite{schwinger1961brownian}, TFD is particularly effective when considering low orders of the scattering matrix. Several studies have successfully applied this theory \cite{alesandrogravitacional, Casimir, santos2017lorentz, santos2016quantized, cabral2023thermal, cabral2024lorentz}. The primary advantage of the TFD formalism is its ability to incorporate temporal information alongside thermal effects, providing a comprehensive framework for studying such interactions. Furthermore, the real time propagators consist of a sum of two parts: one corresponding to the propagator at zero temperature and the other corresponding to a temperature dependent part. Using this formalism the thermal cross section for gravitational Compton scattering can be calculated.

This paper is organized as follows. In Section \ref{II}, the GEM theory and its Lagrangian formalism are presented. In Section \ref{III}, gravitational Compton scattering at zero temperature is investigated. The cross section is calculated, and some discussions are developed. In section \ref{IV}, the TFD formalism is introduced. Then the gravitational Compton scattering at finite temperature is considered. The differential cross section at finite temperature is determined. Some limits of temperature are analyzed. In section \ref{V}, some concluding remarks are made.

\section{Gravitoelectromagnetism (GEM)}\label{II}

In this section, a brief introduction to gravitoelectromagnetism (GEM) is presented. The GEM theory is a gravitational theory constructed by following the similarities between the electromagnetic force and the gravitational force. This gravitational model can be built based on three different situations: (i) considering the similarity between the linearized Einstein and Maxwell equations \cite{Mashhon}; (ii) using tidal tensors \cite{Filipe}; and (iii) decomposing the  Weyl tensor  ($C_{ijkl}$) into $\mathbf{B}_{ij}$ and $\mathbf{E}_{ij}$, representing the gravitomagnetic and gravitoelectric components, respectively \cite{Maartens}. It is important to note that all the aforementioned situations are interrelated. The structure of each framework varies depending on the specific physical phenomena being investigated. Here, the third approach is considered. Thus, the GEM theory is based on the decomposition of the Weyl tensor, which is defined as
\bea
C_{\alpha\sigma\mu\nu}&=&R_{\alpha\sigma\mu\nu}-\frac{1}{2}\left(R_{\nu\alpha}g_{\mu\sigma}+R_{\mu\sigma}g_{\nu\alpha}-R_{\nu\sigma}g_{\mu\alpha}-R_{\mu\alpha}g_{\nu\sigma}\right)\nonumber\\
&+&\frac{1}{6}R\left(g_{\nu\alpha}g_{\mu\sigma}-g_{\nu\sigma}g_{\mu\alpha}\right),
\eea
where $R_{\alpha\sigma\mu\nu}$ is the Riemann tensor, $R_{\mu\nu}$ is the Ricci tensor and $R$ is the Ricci scalar. Using this quantities, the gravitational fields are written as
\bea
\mathbf{B}_{ij}&=&\frac{1}{2}\epsilon_{ikl}C^{kl}_{0j},\\
\mathbf{E}_{ij}&=&-C_{0i0j}.
\eea
 The Weyl tensor shares the same symmetry as the Riemann tensor and has invariance under conformal changes in the metric \cite{danehkar2009significance}. In this context, the GEM, or Maxwell-like,  equations are given as
\begin{eqnarray}
    \partial^{i}\mathbf{E}^{ij}&=&4\pi G\rho^j,\label{2.2}\\
    \partial^{i}\mathbf{B}^{ij}&=&0,\label{2.3}\\
    \varepsilon^{\langle ikl}\partial^k \mathbf{B}^{lj\rangle}-\frac{1}{c}\frac{\partial \mathbf{E}^{ij}}{\partial t}&=&\frac{4\pi G}{c}\mathbf{J}^{ij},\label{2.4}\\
    \varepsilon^{\langle ikl}\partial^k \mathbf{E}^{lj\rangle}+\frac{1}{c}\frac{\partial \mathbf{B}^{ij}}{\partial t}&=&0,\label{2.5}
\end{eqnarray}
where $\mathbf{E}^{ij}$ and $\mathbf{B}^{ij}$ are traceless second-order symmetric tensors, $\rho^j$ is the mass density vector, $\mathbf{J}^{ij}$ is a traceless second-order tensor representing the mass current density, and $G$ is the gravitational constant. Here, the symbol $\langle i\cdots j \rangle$ denotes the
symmetrization of the first and last indices.

To investigate gravitational applications using the GEM theory, the Lagrangian formulation has been developed \cite{Jair}. To achieve this objective, a gravitoelectromagnetic tensor potential $A^{\mu\nu}$ is defined. From this gravitational tensor potential and considering the GEM counterpart of the electromagnetic scalar potential $\varphi$, the GEM fields can be written as
\begin{eqnarray}
    \mathbf{E}&=&-grad\;\varphi-\frac{1}{c}\frac{\partial\widetilde{A}}{\partial t}\nonumber \\ 
    \mathbf{B}&=&curl\;\widetilde{A},
\end{eqnarray}
where $\Tilde{A}$ has components $A^{ij}$, with $i, j = 1, 2, 3$. From this, we can define the gravitoelectromagnetic tensor $F^{\mu\nu\rho}$  as
\begin{eqnarray}
\mathbf{F}^{\mu\nu\rho}=\partial^\mu A^{\nu\rho}-\partial^\nu A^{\mu\rho},\label{GEMFTENSOR}
\end{eqnarray}
where $\mu,\nu,\rho=0,1,2,3$. Thus, we can rewrite the Maxwell-like equations as
\begin{eqnarray}
\partial_{\mu} \mathbf{F}^{\mu\nu\rho}&=&4\pi G\mathcal{J}^{\nu\rho},\label{10}\label{GEMFIELDEQ}\\
\partial_\mu \mathbf{G}^{\mu\langle\nu\rho\rangle}&=&0,\label{11}
\end{eqnarray}
where $\mathcal{J}^{\nu\rho}$ is the mass density tensor depending on $\rho^{j}$ and $\mathbf{J}^{ij}$. Additionally, $\mathbf{G}^{\mu\nu\rho}$  is the dual tensor of GEM, defined as follows
\begin{eqnarray}
\mathbf{G}^{\mu\nu\rho}=\frac{1}{2}\epsilon^{\mu\nu\gamma\sigma} \eta^{\rho\lambda} \mathbf{F}_{\gamma\sigma\lambda}.
\end{eqnarray}

With these ingredients, the GEM Lagrangian is defined as
\begin{eqnarray}
     \mathcal{L}_{\text{GEM}}=-\frac{1}{16\pi}\mathbf{F}_{\mu\nu\rho}\mathbf{F}^{\mu\nu\rho}-G\mathcal{J}^{\nu\rho} A_{\nu\rho}.\label{13}
\end{eqnarray}
Using this Lagrangian in the Euler-Lagrange equations recovers the GEM equations (\ref{10}) and (\ref{11}). Furthermore, this Lagrangian formulation allows the study of interactions between gravitons and other fundamental particles. Moreover, it is important to note that this Lagrangian formulation can be connected to GR in the weak-field limit \cite{alesandrogravitacional}. By substituting Eq. (\ref{GEMFTENSOR}) into Eq. (\ref{GEMFIELDEQ}), we obtain 
	\begin{eqnarray}
		\Box A^{\nu\rho}-\partial^\nu(\partial_\mu A^{\mu\rho})=4\pi G\mathcal{J}^{\nu\rho},
	\end{eqnarray}  
	which allows us to impose the Lorentz-like gauge condition, $\partial_\mu A^{\mu\rho}=0$, due to the gauge freedom of the GEM potentials. Under this condition, the equation simplifies to
	\begin{eqnarray}
		\Box A^{\nu\rho}=4\pi G\mathcal{J}^{\nu\rho}.\label{Aequation}
	\end{eqnarray}  
This result can be directly compared to the linearized Einstein field equation in terms of the metric perturbation \( h_{\mu\nu} \) in General Relativity
\begin{eqnarray}
	\Box \bar{h}_{\mu\nu}=16\pi G T_{\mu\nu},\label{hpertubation}
\end{eqnarray}  
where \( \bar{h}_{\mu\nu}=h_{\mu\nu}-\frac{1}{2}\eta_{\mu\nu} h \) \cite{misner1973k}, with \( h \) being the trace of \( h_{\mu\nu} \). Thus, Eqs. (\ref{Aequation}) and (\ref{hpertubation}) illustrate the formal similarity between GEM theory and the linearized formulation of General Relativity in the weak-field limit.

Since our main objective is to investigate the interaction between gravitons and fermions via gravitational Compton scattering, the complete Lagrangian that describes gravitons, fermions, and their interaction is given as
\bea
\mathcal{L}_{Total}= \mathcal{L}_{\text{G}}+ \mathcal{L}_{\text{F}}+ \mathcal{L}_{\text{I}},
\eea
where
\begin{eqnarray}
     \mathcal{L}_{\text{G}}=-\frac{1}{16\pi}\mathbf{F}_{\mu\nu\rho}\mathbf{F}^{\mu\nu\rho},
\end{eqnarray}
is the free part of the GEM Lagrangian,
\bea
\mathcal{L}_{\text{F}}=-\frac{i}{2}(\bar{\psi}\gamma^\mu\partial_\mu\psi-\partial_\mu\bar{\psi}\gamma^\mu\psi)+m\bar{\psi}\psi
\eea
describes the fermions $\psi$ with mass $m$ and 
\bea
 \mathcal{L}_{\text{I}}=-\frac{i\kappa}{4}A_{\mu\nu}(\Bar{\psi}\gamma^\mu\partial^\nu\psi-\partial^\mu\Bar{\psi}\gamma^\nu\psi),\label{lagrangeint}
\eea
is the interaction part, with $\kappa=\sqrt{8\pi G}$ being the coupling constant. 

From the interaction Lagrangian, we can investigate gravitational Compton scattering, which describes the interactions between gravitons and fermions. The main objective is to calculate the cross-section for this scattering at zero and finite temperature. In the next section, the cross-section for gravitational Compton scattering at zero temperature is determined.

\section{Gravitational Compton scattering at zero temperature}\label{III}

\begin{figure}[h]
    \centering
    \includegraphics[scale=0.3]{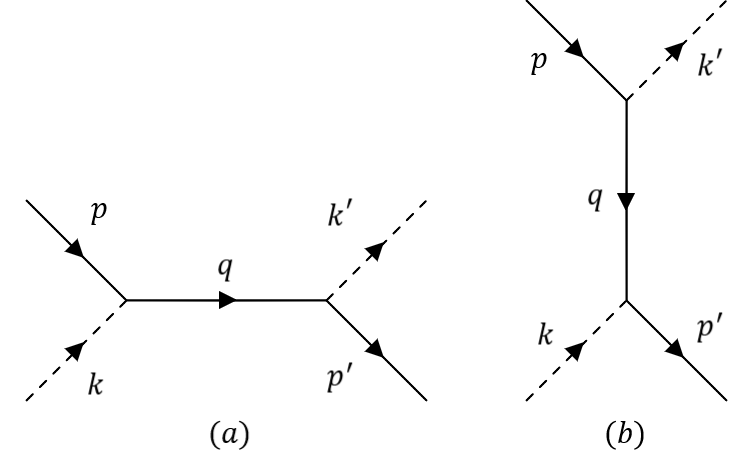}
    \caption{Feynman diagram for gravitational Compton scattering. The diagrams (a) and (b) represent the s-channel and u-channel, respectively. Here, $p_i = p$ and $k_i = k$ represent the initial momenta of the fermion and graviton, respectively, while $p_f = p'$ and $k_f = k'$ represent the final momenta. And $q$ is the momentum of the fermion in the propagator.}
    \label{fig1}
\end{figure}

In this section, we will study the interaction between gravitons and fermions through Compton scattering. This investigation is conducted without considering temperature effects. This scattering process consists of $e^-+g\rightarrow e^-+g$, where $e^-$ and $g$ represent a electron and a graviton, respectively. Figure \ref{fig1} shows the Feynman diagram that describes this process.

In order to calculate the cross section for this scattering process, let's first determine the transition amplitude, which is defined as
\begin{eqnarray}
    \mathcal{M}=\bra{f}\hat{S}^{(2)}\ket{i},\label{18}
\end{eqnarray}
where $\hat{S}^{(2)}$ is the second order scattering matrix given as
\begin{eqnarray}
     \hat{S}^{(2)}=-\frac{1}{2}\int\int d^4x\;d^4y \;\tau[\mathcal{L}_I(x)\mathcal{L}_I(y)],
\end{eqnarray}
with $\tau$ being the time-ordering operator. The final and initial asymptotic states of the particles,  $\bra{f}$ and $\ket{i}$, are written as
\begin{eqnarray}
    \ket{i}&=&\ket{e_i,g_i}=b_{s,p}^\dagger a_{\lambda,k}^\dagger\ket{0},\nonumber\\
     \ket{f}&=&\ket{e_f,g_f}=b_{s^\prime,p^\prime}^\dagger a_{\lambda^\prime,k^\prime}^\dagger\ket{0},
\end{eqnarray}
where $e_i$, $e_f$, $g_i$ and $g_f$ represent the initial and final states of the electron and graviton, respectively. Here, $a^\dagger$ is the graviton creation operator  and $b^\dagger$ is the electron creation operator.

Using the interaction Lagrangian (\ref{lagrangeint}), the transition amplitude is written as
\begin{eqnarray}
    \mathcal{M}&=&\frac{\kappa^2}{16}\int d^4x\;d^4y\bra{f}\tau\big[A_{\mu\nu}A_{\alpha\beta}(\Bar{\psi}_x\gamma^{\mu}\partial^{\nu}\psi_x-\partial^{\mu}\Bar{\psi}_x\gamma^\nu \psi_x)\nonumber\\
    &&\times(\Bar{\psi}_y\gamma^{\alpha}\partial^{\beta}\psi_y-\partial^{\alpha}\Bar{\psi}_y\gamma^\beta \psi_y)\big]\ket{i}.
\end{eqnarray}
The only terms that will contribute to the process are those represented by the Feynman diagrams. Therefore, after analyzing the operators contained in $\mathcal{M}$, we can state that the probability amplitude becomes
\begin{eqnarray}
    \mathcal{M}&=&\frac{\kappa^2}{16}\int d^4x\;d^4y\bra{f}\tau\big[\big(A_{\mu\nu}^-A_{\alpha\beta}^+ +A_{\mu\nu}^+A_{\alpha\beta}^-\big)\nonumber\\
    &&\times\big(\Bar{\psi}^-_x\gamma^{\mu}[\partial^\nu\psi_x\Bar{\psi}_y]\gamma^{\alpha}\partial^{\beta}\psi^+_y-\Bar{\psi}^-_x\gamma^\mu[\partial^\nu\psi_x\partial^\alpha\Bar{\psi}_y]\gamma^\beta\psi_y^+\nonumber\\
    &&-\partial^\mu\Bar{\psi}^-_x\gamma^\nu[\psi_x\Bar{\psi}_y]\gamma^\alpha\partial^\beta\psi^+_y+\partial^\mu\Bar{\psi}^-_x\gamma^\nu[\psi_x\partial^\alpha\Bar{\psi}_y]\gamma^\beta\psi_y^+\big)\big]\ket{i},
\end{eqnarray}
where negative and positive indices represent the parts of the fields associated with their respective frequencies. In other words, for the fields of GEM, we have that
\begin{eqnarray}
    &&A_{\mu\nu}^-=\int d^3k\;N_k\sum_\lambda\epsilon_{\mu\nu}^{(\lambda)}a^\dagger_{\lambda,k}\;e^{ikx}\nonumber\\
    &&A_{\mu\nu}^+=\int d^3k\;N_k\sum_\lambda\epsilon_{\mu\nu}^{(\lambda)}a_{\lambda,k}\;e^{-ikx}\label{AGEM}
\end{eqnarray}
with $N_k$ being the normalization constant and $\epsilon_{\mu\nu}$ the graviton polarization tensor. For the fermionic field, on the other hand, we will have that
\begin{eqnarray}
    &&\Bar{\psi}_x^-=\int d^3p\;N_p \sum_s\Bar{u}^{(s)}_p b_{s,p}^\dagger\;e^{
ipx}\nonumber\\
&&\psi_x^+=\int d^3p\;N_p \sum_s u^{(s)}_p b_{s,p}\;e^{-ipx}.\label{FERMIDIRAC}
\end{eqnarray}
Here, $N_p$ is a normalization constant and $u^{(s)}_p$ is the Dirac spinor. With these ingredients, the transition amplitude can be written as
\begin{eqnarray}
     \mathcal{M}=\mathcal{M}_{I}+\mathcal{M}_{II},\label{25}
\end{eqnarray}
where $\mathcal{M}_{I}$ and $\mathcal{M}_{II}$ represent Feynman diagrams (a) and (b) from Figure \ref{fig1}, respectively, and are given as
\begin{eqnarray}
    \mathcal{M}_{I}=&-&\frac{\kappa^2}{16}\frac{1}{(k_f+p_f)^2-m^2}\sum_{\lambda s}\epsilon_{\mu\nu}^{(\lambda_f)}\epsilon_{\alpha\beta}^{(\lambda_i)}\big[\Bar{u}^{(s_f)}_{p_f}\gamma^\mu(k_f^\nu+p_f^\nu)(\slashed{k}_f+\slashed{p}_f+m)\gamma^\alpha p_i^\beta u^{(s_i)}_{p_i}\nonumber\\
    &+&\Bar{u}^{(s_f)}_{p_f}\gamma^\mu(k_f^\nu+p_f^\nu)(\slashed{k}_f+\slashed{p}_f+m)
    (k_f^\alpha+p_f^\alpha) \gamma^\beta u^{(s_i)}_{p_i}+p_f^\mu\Bar{u}^{(s_f)}_{p_f}\gamma^\nu(\slashed{k}_f+\slashed{p}_f+m)\gamma^\alpha p_i^\beta u^{(s_i)}_{p_i}\nonumber\\
    &+&p_f^\mu \Bar{u}^{(s_f)}_{p_f}\gamma^\nu(\slashed{k}_f+\slashed{p}_f+m)(k_f^\alpha +p_f^\alpha)\gamma^\beta u^{(s_i)}_{p_i}\big]\label{M1}
\end{eqnarray}
and
\begin{eqnarray}
    \mathcal{M}_{II}=&-&\frac{\kappa^2}{16}\frac{1}{(p_i-k_f)^2-m^2}\sum_{\lambda s}\epsilon_{\mu\nu}^{(\lambda_f)}\epsilon_{\alpha\beta}^{(\lambda_i)}\big[\Bar{u}^{(s_f)}_{p_f}\gamma^\mu(p_f^\nu-k_i^\nu)(\slashed{p}_i-\slashed{k}_f+m)\gamma^\alpha p_i^\beta u^{(s_i)}_{p_i}\nonumber\\
    &+&\Bar{u}^{(s_f)}_{p_f}\gamma^\mu(p_i^\nu-k_f^\nu)(\slashed{p}_i-\slashed{k}_f+m)(p_f^\alpha-k_i^\alpha)\gamma^\beta u^{(s_i)}_{p_i}+p_f^\mu \Bar{u}^{(s_f)}_{p_f}\gamma^\nu(\slashed{p}_i-\slashed{k}_f+m)\gamma^\alpha p_i^\beta u^{(s_i)}_{p_i}\nonumber\\
    &+&p_f^\mu \Bar{u}^{(s_f)}_{p_f}\gamma^\nu(\slashed{p}_i-\slashed{k}_f+m)(p_i^\alpha -k_f^\alpha)\gamma^\beta u^{(s_i)}_{p_i}\big],\label{M2}
\end{eqnarray}
where we use the fact that the fermion propagator is defined by
\begin{eqnarray}
    \bra{0}\tau[\psi_x \Bar{\psi}_y]\ket{0}=\int \frac{d^4q}{(2\pi)^4}e^{-iq(x-y)}\frac{\slashed{q}+m}{q^2-m^2},
\end{eqnarray}
with $q$ being the fermion momentum. After some calculations, the total transition amplitude (\ref{25}) becomes
\begin{eqnarray}
    \mathcal{M}=&-&\frac{\kappa^2}{16}\frac{1}{(k_f+p_f)^2-m^2}\sum_{\lambda s}[\Bar{u}^{(s_f)}_{p_f}(2p_f+k_f)\cdot\epsilon_f^*\slashed{\epsilon}_f^*](\slashed{k}_f+\slashed{p}_f+m)\nonumber\\
    &\times&[\slashed{\epsilon}_i\epsilon_i\cdot(2p_i+k_i) u^{(s_i)}_{p_i}]\nonumber\\
    &-&\frac{\kappa^2}{16}\frac{1}{(p_i-k_f)^2-m^2}\sum_{\lambda s}[\Bar{u}^{(s_f)}_{p_f}(2p_f-k_i)\cdot\epsilon_i\slashed{\epsilon}_i](\slashed{p}_i-\slashed{k}_f+m)\nonumber\\
    &\times&[\slashed{\epsilon}_i^*\epsilon_i^*\cdot(2p_i-k_f)u^{(s_i)}_{p_i}],
\end{eqnarray}
where have been used that $\epsilon_{\mu\nu}=\epsilon_\mu \epsilon_\nu$ and $\slashed{\epsilon}=\gamma^\mu \epsilon_\mu$. 

With the transition amplitude, we can calculate the cross section for gravitational Compton scattering at zero temperature. This approach will be carried out considering the center-of-mass (CM) reference frame, where the momenta are given by
\begin{eqnarray}
    p_i&=&(\omega,0,0,\omega),\nonumber\\
    k_i&=&(\omega,0,0,-\omega),\nonumber\\
    p_f&=&(\omega,-\omega\sin{\theta},0,-\omega\cos{\theta}),\nonumber\\
    k_f&=&(\omega,\omega\sin{\theta,0,\omega\cos{\theta}}),
\end{eqnarray}
where $\omega$ represents the frequency and $\theta$ the scattering angle. It is important to note that, due to the graviton's mass being zero, $m_g=0$, it follows from the dispersion relation that $|E_k|^2=|\vec{p}|^2=|\vec{p^\prime}|^2$. Furthermore, by definition, the cross section is expressed as
\begin{eqnarray}
    \Big(\frac{d\sigma}{d\Omega}\Big)=\frac{1}{64\pi^2 s}\langle|\mathcal{M}|^2\rangle,\label{3.2}
\end{eqnarray}
with $s=(2\omega)^2$. The parameter $\langle|\mathcal{M}|^2\rangle$  represents the probability density, expressed in terms of the modulus of the transition amplitude, that is,
\begin{eqnarray}
    \langle|\mathcal{M}|^2\rangle=\frac{1}{4}\sum_s|\mathcal{M}|^2=\langle|\mathcal{M}_I|^2\rangle+\langle|\mathcal{M}_{II}|^2\rangle+\langle2\Re\mathcal{M}_I^\dagger\mathcal{M}_{II}\rangle.\label{3.3}
\end{eqnarray}

Now let's calculate these terms separately. The first term is given as
\begin{eqnarray}
    \langle|\mathcal{M}_I|^2\rangle&=&\frac{1}{4}\sum_\lambda\sum_s\Big(\frac{\kappa^2}{16}\frac{1}{(k_f+p_f)^2-m^2}\Big)^2[(2p_f^\nu+k_f^\nu)\epsilon_\nu^*\epsilon_\mu^*][(2p_i^\beta+k_i^\beta)\epsilon_\beta\epsilon_\alpha]\nonumber\\
    &\times&[(2p_i^\sigma+k_i^\sigma)\epsilon_\sigma^*\epsilon_\rho^*][(2p_f^\xi+k_f^\xi)\epsilon_\xi\epsilon_\lambda][\Bar{u}^{(s_f)}_{p_f}\gamma^\mu(\slashed{k}_f+\slashed{p}_f+m)\gamma^\alpha u^{(s_i)}_{p_i}\nonumber\\
    &\times&\Bar{u}^{(s_i)}_{p_i}\gamma^\rho(\slashed{k}_f+\slashed{p}_f+m)\gamma^\lambda u^{(s_f)}_{p_f}].\label{33}
\end{eqnarray}
Using the property of the trace of gamma matrices
\begin{eqnarray}
    &&\sum_s[\Bar{u}^{(s_f)}_{p_f}\gamma^\mu(\slashed{k}_f+\slashed{p}_f+m)\gamma^\alpha u^{(s_i)}_{p_i}\Bar{u}^{(s_i)}_{p_i}\gamma^\rho(\slashed{k}_f+\slashed{p}_f+m)\gamma^\lambda u^{(s_f)}_{p_f}]\nonumber\\
    &&=\mathrm{Tr}[\sum_su^{(s_f)}_{p_f}\Bar{u}^{(s_f)}_{p_f}\gamma^\mu(\slashed{k}_f+\slashed{p}_f+m)\gamma^\alpha u^{(s_i)}_{p_i}\Bar{u}^{(s_i)}_{p_i}\gamma^\rho(\slashed{k}_f+\slashed{p}_f+m)\gamma^\lambda ],\label{trace}
\end{eqnarray}
the completeness relation
\begin{eqnarray}
    \sum_su(p,s)\Bar{u}(p,s)&=&\slashed{p}_i+m,\label{2.28}
\end{eqnarray}
and the summation over the polarization tensor
\bea
 \sum_\lambda\epsilon_{\mu\nu}(k,\lambda)\epsilon_{\alpha\rho}(k,\lambda)&=&\frac{1}{2}(g_{\mu\nu}g_{\nu\rho}+g_{\mu\rho}g_{\nu\alpha}-g_{\mu\nu}g_{\alpha\rho}),
\eea
Eq. (\ref{33}) becomes
\begin{eqnarray}   \langle|\mathcal{M}_I|^2\rangle&=&\frac{\kappa ^4}{4096 \left((k_f+p_f)^2-m^2\right)^2}\left({k_f}^{\nu }+2 {p_f}^{\nu }\right) \left({k_f}^{\xi }+2 {p_f}^{\xi }\right)
   \left({k_i}^{\beta }+2 {p_i }^{\beta }\right) \left({k_i}^{\sigma }+2 {p_i }^{\sigma }\right)\nonumber\\
   &\times&\left({g}^{\alpha \sigma }
   {g}^{\beta \rho }+{g}^{\alpha \rho } {g}^{\beta \sigma }-{g}^{\alpha \beta } {g}^{\rho \sigma }\right) \left(-{g}^{\lambda \xi } {g}^{\mu \nu
   }+{g}^{\lambda \nu } {g}^{\mu \xi }+{g}^{\lambda \mu } {g}^{\nu \xi }\right)\nonumber\\
   &\times&\mathrm{Tr}[\left({\gamma }\cdot {p_f}+m\right){\gamma }^{\mu }\left({\gamma }\cdot
   \left({k_f}+{p_f}\right)+m\right){\gamma }^{\alpha }\left({\gamma }\cdot p_i+m\right){\gamma }^{\rho }\left({\gamma }\cdot\left({k_f}+{p_f}\right)+m\right){\gamma }^{\lambda }].\nonumber\\
\end{eqnarray}
Introducing the Mandelstam variables, defined by
\begin{eqnarray}
    s=2p_i\cdot k_i=2p_f\cdot k_f;\quad t=-2p_i\cdot p_f=-2k_i\cdot k_f;\quad u=-2p_i\cdot k_f=-2p_f\cdot k_i,
\end{eqnarray}
or, in terms of the scattering angle $\theta$,
\begin{eqnarray}
    s=(2\omega)^2; \quad t=-2\omega^2(1-\cos{\theta});\quad u=-2\omega^2(1+\cos{\theta}),
\end{eqnarray}
we obtain
\begin{eqnarray}
    \langle|\mathcal{M}_I|^2\rangle=-\frac{1}{16} \kappa ^4 \omega ^4 (\cos {\theta }-1).
\end{eqnarray}
Similarly, we find that
\begin{eqnarray}
    \langle|\mathcal{M}_{II}|^2\rangle&=&-\frac{1}{16} \kappa ^4 \omega ^4 (\cos {\theta }-1),\\
     \langle2\Re\mathcal{M}_I^\dagger\mathcal{M}_{II}\rangle&=&\frac{1}{128} \kappa ^4 \omega ^4 (28\cos {\theta }+3 \cos {2\theta }-7).
\end{eqnarray}

Therefore, the total transition amplitude is given as
\begin{eqnarray}
    \langle|\mathcal{M}|^2\rangle=\frac{3}{16} \kappa ^4 \omega ^4 \cos ^4\left(\frac{\theta }{2}\right).
\end{eqnarray}
Thus, the differential cross section is obtained as
\begin{eqnarray}
    \Big(\frac{d\sigma}{d\Omega}\Big)&=&\frac{3 \kappa ^4 \omega ^2 \cos ^4\left(\frac{\theta }{2}\right)}{4096 \pi ^2}.\label{44}
\end{eqnarray}
Performing the integration over the solid angle, the cross section of the gravitational Compton scattering is
\begin{eqnarray}
     \sigma&=&\frac{\kappa^4\omega^2}{1024\pi}.
\end{eqnarray}
Although we are using a gravitational formalism similar to that of electromagnetism, there are fundamental differences between the two. While in electromagnetism charges generate the field, in GEM, the field is generated by masses. Additionally, in electromagnetism, the field is described by vectors, while in GEM it is described by tensors. Furthermore, the coupling constants are different. From \cite{cabral}, we have that the cross section for the Compton scattering is given by
\begin{eqnarray}
		\bigg(\frac{d\sigma}{d\Omega}\bigg)_{\text{QED}}=\frac{\alpha^2}{2(2\omega)^2}\bigg(\frac{\cos\theta+1}{2}+\frac{2}{\cos\theta+1}\bigg).
\end{eqnarray}
Note that the coupling constant $\alpha$, which is the fine-structure constant $(\alpha = e^2 / 4\pi\epsilon_0)$, is dimensionless, whereas the gravitational coupling constant $\kappa$ has the dimension of energy $(\kappa \approx \sqrt{G} \approx 1/M_P)$.  Therefore, to compare the cross section of GEM with the QED result, the GEM result must be multiplied by a characteristic energy scale to match the units of the QED cross section. This is achieved by defining $\kappa \to \kappa^\prime = \kappa E_c$, where $E_c$ is a scattering energy scale. These distinctions imply that the cross section for gravitational Compton scattering may exhibit significant differences between the two theories.

In the next section, we will analyze the thermal effects in this gravitational scattering process.

\section{Gravitational Compton scattering at finite temperature} \label{IV}

The main objective here is to incorporate the effects of temperature into the cross section of gravitational Compton scattering. The thermal effects are introduced using the TFD formalism. This is a real-time formalism that introduces temperature effects into a system without losing temporal information \cite{Umezawa1, Umezawa2, Book, Umezawa22, Khanna1, Khanna2, Santana1, Santana2}. TFD is characterized by two ingredients: (1) duplication of the Hilbert space and (2) the Bogoliubov transformation. The doubled Hilbert space, or thermal Hilbert space, is defined as $\mathcal{S}_T = \mathcal{S} \otimes \tilde{\mathcal{S}}$, where $\mathcal{S}$ is the standard Hilbert space and $\tilde{\mathcal{S}}$ is the dual or tilde Hilbert space. Although the duplication of the Hilbert space for the introduction of thermal effects might suggest the presence of duplicated quantities and terms, which indeed occurs mathematically, it is important to note that the physical quantities we aim to measure are represented in the usual space $\mathcal{S}$. Therefore, the duplication of the space does not lead to any unphysical or absurd consequences in the results. Instead, it is merely a method to introduce thermal effects, allowing us to later select only the elements of the usual Hilbert space \cite{Book, das2023finite}. Moreover, TFD not only introduces temperature into the fields through the duplication of the Hilbert space and the Bogoliubov transformation (which we will discuss next), but also quantizes the quantities associated with thermal effects. These considerations enable us to view TFD as a process of thermal quantization. With that being said, the relationship between the two fundamental ingredients mentioned earlier is governed by the tilde conjugation rules, which are defined as follows:
\begin{eqnarray}
    &&(A_iA_j)^\til=\Tilde{A}_i\Tilde{A}_j;\nonumber\\
    &&(cA_i+A_j)^\til=c^*\Tilde{A}_i+\Tilde{A}_j;\nonumber\\
    &&(A_i^\dagger)^\til=(\Tilde{A}_i)^\dagger;\\
    &&(\Tilde{A}_i)^\til=\pm A_i;\nonumber\\
   &&\comm{A_i}{\Tilde{A}_j}=0.\nonumber\label{operator}
\end{eqnarray}	

The Bogoliubov transformation introduces temperature effects through a rotation between the tilde and non-tilde variables. As an example, let's consider the transformations for fermions operators
\begin{eqnarray}
    b_{s,p}=U(\beta)b_{s,p}(\beta)+V(\beta)\Tilde{b}_{s,p}^\dagger(\beta),\nonumber\\
    \Tilde{b}_{s,p}=U(\beta)\Tilde{b}_{s,p}(\beta)-V(\beta)b_{s,p}^\dagger(\beta),\nonumber\\
    b_{s,p}^\dagger=U(\beta)b_{s,p}^\dagger(\beta)+V(\beta)\Tilde{b}_{s,p}(\beta),\nonumber\\
    \Tilde{b}_{s,p}^\dagger=U(\beta)\Tilde{b}_{s,p}^\dagger(\beta)-V(\beta)b_{s,p}(\beta),\label{boguferm}
\end{eqnarray}
where $U(\beta)=\cos{\theta}(\beta)$ and $V(\beta)=\sin{\theta}(\beta)$. In a similar way, for bosons we have 
\begin{eqnarray}
a_{\lambda,k}=U^\prime(\beta)a_{\lambda,k}(\beta)+V^\prime(\beta)\Tilde{a}_{\lambda,k}^\dagger(\beta),\nonumber\\
    \Tilde{a}_{\lambda,k}=U^\prime(\beta)\Tilde{a}_{\lambda,k}(\beta)+V^\prime(\beta)a_{\lambda,k}^\dagger(\beta),\nonumber\\
    a_{\lambda,k}^\dagger=U^\prime(\beta)a_{\lambda,k}^\dagger(\beta)+V^\prime(\beta)\Tilde{a}_{\lambda,k}(\beta),\nonumber\\
    \Tilde{a}_{\lambda,k}^\dagger=U^\prime(\beta)\Tilde{a}_{\lambda,k}^\dagger(\beta)+V^\prime(\beta)a_{\lambda,k}(\beta),\label{bogubos}
\end{eqnarray}
with $U^\prime(\beta)=\cosh{\theta}(\beta)$ and $V^\prime(\beta)=\sinh{\theta}(\beta)$. Note that the operators shown in Eqs. (\ref{boguferm}) and (\ref{bogubos}) are the same as those in Eqs. (\ref{AGEM}) and (\ref{FERMIDIRAC}), respectively.

Another quantity that will be modified in the TFD formalism is the propagator, which will be decomposed into two distinct parts: one representing the spatial contribution and the other describing the thermal effects. Therefore, it will be written as
\begin{eqnarray}
    \bra{0(\beta)}\tau[\psi_x\Bar{\psi}_y]\ket{0(\beta)}=i\int\frac{d^4q}{(2\pi)^4}e^{-iq(x-y)}\Delta(q,\beta),
\end{eqnarray}
where $\Delta(q,\beta)=S^{(0)}(q)+S^{(\beta)}(q)$ with
\begin{eqnarray}
    S^{(0)}(q)&=&\frac{\slashed{q}+m}{q^2-m^2}
\end{eqnarray}
being the fermion propagator at zero temperature, and
\bea
S^{(\beta)}(q)&=&\frac{2\pi i}{e^{\beta q_0}+1}\Bigg[\frac{(\gamma^0\xi-\gamma\cdot\Vec{q}+m)}{2\xi}\Delta_1\delta(q^0-\xi)+\frac{(\gamma^0\xi+\gamma\cdot\Vec{q}+m)}{2\xi}\Delta_2\delta(q^0+\xi)\Bigg]\label{3.32}
\eea
are the corrections for the propagator due to the temperature effects.
Here, have been used that
\bea
    \Delta_1=\begin{pmatrix}
1 & e^{\beta q_0/2}\\
e^{\beta q_0/2} & -1 \\
\end{pmatrix}, \quad\quad
    \Delta_2=\begin{pmatrix}
-1 & e^{\beta q_0/2}\\
e^{\beta q_0/2} & 1 \\
\end{pmatrix}.
\eea

Now we can calculate the transition amplitude for the gravitational Compton scattering at finite temperature. In this approach, Eq. (\ref{18}) becomes
\begin{eqnarray}
    \mathcal{M}(\beta)=\bra{f,\beta}\hat{S}^{(2)}\ket{i,\beta},
\end{eqnarray}
where the thermal asymptotic states are given by
\begin{eqnarray}
    \ket{f,\beta}&=&b^\dagger_f(\beta)a^\dagger_f(\beta)\ket{0(\beta)},\nonumber\\
    \ket{i,\beta}&=&b^\dagger_i(\beta)a^\dagger_i(\beta)\ket{0(\beta)}.
\end{eqnarray}
The second-order term of the scattering matrix  will be
\begin{eqnarray}
    \hat{S}^{(2)}=-\frac{1}{2}\int d^4x\;d^4y\;\tau[\hat{\mathcal{L}}_I(x)\hat{\mathcal{L}}_I(y)],
\end{eqnarray}
with the interaction Lagrangian for the doubled space, representing the graviton-fermion interaction, being presented in the form
\begin{eqnarray}
    \hat{\mathcal{L}}_I(x)&=&\mathcal{L}_I(x)-\Tilde{\mathcal{L}}_I(x)\label{3.4}\nonumber\\
    &=&-\frac{i\kappa}{4}A_{\mu\nu}(\Bar{\psi}_x\gamma^\mu\partial^\nu \psi_x-\partial^\mu\Bar{\psi}_x\gamma^\nu\psi_x)+\frac{i\kappa}{4}\Tilde{A}_{\mu\nu}(\Tilde{\Bar{\psi}}_x\gamma^\mu\partial^\nu \Tilde{\psi}_x-\partial^\mu\Tilde{\Bar{\psi}}_x\gamma^\nu\Tilde{\psi}_x).\label{LI}
\end{eqnarray}
Then the transition amplitude becomes
\begin{eqnarray}
\mathcal{M}(\beta)=\mathcal{M}_{\RN{1}}(\beta)+\mathcal{M}_{\RN{2}}(\beta).
\end{eqnarray}

The procedure to calculate the transition amplitude at finite temperature is the same as described for the case at zero temperature. Considering only the contributions relevant to the analyzed process, the transition amplitudes at finite temperature corresponding to the Feynman diagrams in Figure \ref{fig1} are given as
\begin{eqnarray}
    \mathcal{M}_{\RN{1}}(\beta)=&-&\frac{i \kappa^2}{16}U^2(\beta)U^{\prime2}(\beta)\sum_{\lambda s}[\Bar{u}_{p_f}^{(s_f)}(2p_f+k_f)\cdot\epsilon_{f}^{*g}\slashed{\epsilon}_{f}^{*g}]\nonumber\\
    &\times&\Delta_{(p_f+k_f)}[\slashed{\epsilon}_i^{g}\epsilon_i^{g}\cdot(2p_i+k_i)u_{p_i}^{(s_i)}],\\
     \mathcal{M}_{\RN{2}}(\beta)=&-&\frac{i\kappa^2}{16}U^2(\beta)U^{\prime2}(\beta)\sum_{\lambda s}[\Bar{u}^{(s_f)}_{p_f}(2p_f-k_i)\cdot\epsilon_i^{g}\slashed{\epsilon}_i^{g}]\nonumber\\
    &\times&\Delta^{ab}_{(p_f-k_i)}[\slashed{\epsilon}_f^{*g}\epsilon_f^{*g}\cdot(2p_i-k_f)u_{p_i}^{(s_i)}].
\end{eqnarray}

To obtain the cross section at finite temperature, we need the quantity
\begin{eqnarray}
     \langle|\mathcal{M}(\beta)|^2\rangle=\frac{1}{4}\sum_s|\mathcal{M}(\beta)|^2=\langle|\mathcal{M_\RN{1}}(\beta)|^2\rangle+\langle|\mathcal{M_\RN{2}}(\beta)|^2\rangle+\langle2\Re\mathcal{M_\RN{1}}^\dagger(\beta)\mathcal{M_\RN{2}}(\beta)\rangle.\label{61}
\end{eqnarray}
Taking the center of mass as the reference frame, we obtain the first term of Eq. (\ref{61}) as
\begin{eqnarray}
  \langle|\mathcal{M_\RN{1}}(\beta)|^2\rangle&=&-\frac{1}{16} \kappa^4 \omega^4 (\cos{\theta} - 1)\frac{1+\mathcal{A}(\beta)}{(1+e^{-2\beta\omega})^2}\bigg(\frac{1+\coth{\beta\omega}}{2}\bigg)^2,\label{62}
\end{eqnarray}
where
\begin{eqnarray}
   \mathcal{A}(\beta)=e^{-2\beta\omega}+e^{-4\beta\omega}\big(4\pi^2\omega^2\Gamma_1^2+1\big),
\end{eqnarray}
with $\Gamma_1=\big(\delta(2\omega)\Delta_1-\delta(2\omega)\Delta_2\big)$. The second term of Eq. (\ref{61}) reads
\begin{eqnarray}
    \langle|\mathcal{M_\RN{2}}(\beta)|^2\rangle&=&-\frac{1}{16} \kappa^4 \omega^4 (\cos{\theta} - 1)\frac{1+\mathcal{B}(\beta)}{(1+e^{-2\beta\omega})^2}\bigg(\frac{1+\coth{\beta\omega}}{2}\bigg)^2,\label{64}
\end{eqnarray}
where
\begin{eqnarray}
   \mathcal{B}(\beta)&=&e^{-2\beta\omega}-e^{-4\beta\omega}\bigg\{\frac{\pi^2\omega^2}{2}\bigg[1+\big(\delta(2\omega-\xi_2)\Delta_1^\prime-3\delta(2\omega+\xi_2)\Delta_2^\prime\big)\Gamma_2\nonumber\\
    &-&\xi_2^{-2}\Big(2\cos\theta\big(\omega^2\Gamma_3^2-2\xi_2^2\delta(2\omega+\xi_2)\Delta_2^\prime\Gamma_2\big)+\cos2\theta\big(\xi_2^2\Gamma_2^2-4\omega^2\Gamma_3^2\nonumber\\
    &-&2\omega^2\cos3\theta\Gamma_3^2+2\xi_2\omega\Gamma_3\Gamma_4\big)+4\omega^2\Gamma_3^2-2\omega\xi_2\Gamma_3\Gamma_4\Big)\bigg]\bigg\},
\end{eqnarray}
with $\Gamma_2=\big(\delta(2\omega-\xi_2)\Delta_1^\prime+\delta(2\omega+\xi_2)\Delta_2^\prime\big)$,\;$\Gamma_3=\big(\delta(2\omega-\xi_2)\Delta_1^\prime-\delta(2\omega+\xi_2)\Delta_2^\prime\big)$ and $\Gamma_4=\big(\delta(2\omega-\xi_2)\Delta_1^\prime+2\delta(2\omega+\xi_2)\Delta_2^\prime\big)$. Finally, the third term is written as
\begin{eqnarray}
     \langle2\Re\mathcal{M_\RN{1}}^\dagger(\beta)\mathcal{M_\RN{2}}(\beta)\rangle&=&\frac{1}{128}\kappa^4\omega^4(28\cos{\theta}+3\cos{2\theta}-7)\frac{1+\eta\;\mathcal{C}(\beta)}{(1+e^{-2\beta\omega})^2}\bigg(\frac{1+\coth{\beta\omega}}{2}\bigg)^2,\;\label{66}
\end{eqnarray}
with
\begin{eqnarray}
    \mathcal{C}(\beta)&=&e^{-2\beta\omega}\big(56\cos\theta+6\cos2\theta-14\big)-e^{-4\beta\omega}\Big\{2\pi^2\omega^2\Big[2\cos\theta\Gamma_1\big(2\delta(2\omega-\xi_2)\Delta_1^\prime\nonumber\\
    &+&17\delta(2\omega+\xi_2)\Delta_2^\prime\big)-\xi_2^{-1}\Big(14\omega\Gamma_1\Gamma_3-45\omega\cos\theta\Gamma_5\Gamma_3+\cos2\theta\Gamma_1\big(21\delta(2\omega-\xi_2)\Delta_1^\prime\nonumber\\
    &+&16\delta(2\omega+\xi_2)\Delta_2^\prime\big)+34\omega\cos2\theta\Gamma_1\Gamma_3+\cos3\theta\Gamma_1\big(3\omega\Gamma_3+2\xi_2\Gamma_2\big) \Big)\nonumber\\
    &+&\Gamma_1\big(9\delta(2\omega-\xi_2)\Delta_1^\prime+16\delta(2\omega+\xi_2)\Delta_2^\prime\big)\Big]+3\xi_2^{-1}-28\cos\theta+7\Big\},
\end{eqnarray}
where $\Gamma_5=\big(\delta(2\omega)\Delta_1+\delta(2\omega)\Delta_2\big)$ and $\eta=1/(28\cos{\theta}+3\cos{2\theta}-7)$. Thus, using Eqs. (\ref{62}), (\ref{64}) and (\ref{66}) the total transition amplitude becomes
\begin{eqnarray}
     \langle|\mathcal{M}(\beta)|^2\rangle&=&\frac{3}{16}\kappa^4\omega^4\cos^4{\Big(\frac{\theta}{2}\Big)}\frac{1-\mathcal{D}(\beta)}{(1+e^{-2\beta\omega})^2}\bigg(\frac{1+\coth{\beta\omega}}{2}\bigg)^2,
\end{eqnarray}
where
\begin{eqnarray}
    \mathcal{D}(\beta)=(\cos\theta-1)(\mathcal{A}(\beta)+\mathcal{B}(\beta))-\frac{\eta\mathcal{C}(\beta)}{8}.
\end{eqnarray}
It is important to note that, in these results, we have $\xi_1=m_g=0$ e $\xi_2=(2\omega^2(1+\sin{\theta}+\cos{\theta})+\omega^2\sin{2\theta})^{1/2}$. 

From these results, our objective is to obtain the differential cross section at finite temperature, which is defined by
\begin{eqnarray}
     \left(\frac{d\sigma}{d\Omega}\right)_\beta&=&\frac{1}{64\pi^2 s}\langle|\mathcal{M}(\beta)|^2\rangle.
\end{eqnarray}
From the total transition amplitude, the differential cross section for the gravitational Compton scattering at finite temperature is given as
\begin{eqnarray}
     \left(\frac{d\sigma}{d\Omega}\right)_\beta &=&\alpha(\beta) \left(\frac{d\sigma}{d\Omega}\right)_{0}\label{71}
\end{eqnarray}
where $\left(\frac{d\sigma}{d\Omega}\right)_{0}$ is the differential cross section at zero temperature as found in Eq. (\ref{44}) and 
\bea
\alpha(\beta)\equiv\frac{1-\mathcal{D}(\beta)}{(1+e^{-2\beta\omega})^2}\bigg(\frac{1+\coth{\beta\omega}}{2}\bigg)^2
\eea
is the modifications due to temperature effects.

It is important to note that our result for the cross section of gravitational Compton scattering is strongly influenced by temperature effects. Two limits must be investigated: (i) the limit of zero temperature and (ii) the limit of very high temperature. First, we note that when the temperature goes to zero, or $\beta\rightarrow \infty$, we have
$\alpha(\beta)\rightarrow 1$
and the differential cross section at zero temperature, Eq. (\ref{44}), is recovered. On the other hand, in the limit of very high temperature, or  $\beta\rightarrow 0$, the function $\coth{\beta\omega}$ becomes very large, and the effect of temperature becomes dominant. Moreover, a comparison of the use of TFD instead of CTP, as shown in \cite{cabral}, reveals that, despite exhibiting similar functional behavior, the results obtained from each formalism are significantly different. This discrepancy arises because the two formalisms are fundamentally distinct. In TFD, we introduce thermal effects through methods like Bogoliubov transformations, such as the doubling of the usual Hilbert space, whereas the CTP approach relies on a different construction. Additionally, the Matsubara method is not applicable to scattering processes, as it does not allow for the definition of transition probabilities, and consequently, the cross-section \cite{xu1995compton}. Given these considerations, the TFD formalism emerges as the optimal choice for studying thermal effects in quantum systems at tree-level scattering \cite{das2023finite}.

Although the study presented here is theoretical, gravitational Compton scattering could conceivably occur in regions of space with high temperatures, such as near black holes or neutron stars. Furthermore, it's important to emphasize that the energies and conditions required to observe such interactions currently exceed our technological capabilities and observational techniques. Therefore, gravitational Compton scattering remains primarily a theoretical concept within the realm of quantum gravity and fundamental physics research.

\section{Conclusions}\label{V}

The Compton scattering process is a fundamental and well-tested phenomenon in QED. To investigate a gravitational version of Compton scattering, the GEM theory is considered. GEM is a weak-field gravitational theory based on the similarities between electromagnetism and gravity. Assuming that the GEM equations are derived from the decomposition of the Weyl tensor, a Lagrangian formulation for this theory has been developed. From this Lagrangian, we can study the interaction between gravitons and other fundamental particles, such as electrons. Here, the gravitational Compton scattering, which describes the interaction between a graviton and an electron, is investigated. Two main results are obtained. First, the cross section at zero temperature is determined. Second, the thermal effects on the cross section are calculated. The temperature effects are introduced using the TFD formalism. Although there are other methods for introducing temperature into quantum systems, such as the Matsubara and CTP methods, TFD is the optimal formalism for studying tree-level processes in thermal equilibrium. Our results at zero temperature show that the gravitational cross section has a structure similar to that of the QED process differing only in the dimensionality of the coupling constants in each theory. Therefore, the GEM coupling constant $\kappa$ must be multiplied by a characteristic energy in order to match the units of the Compton scattering cross section in QED. This difference emphasizes the distinction between the two theories and the unique features of GEM. For the second result, it is evident that the temperature effects alter the cross section. Furthermore, at very high energy temperatures, this effect becomes dominant. Although temperature effects are crucial, it's important to emphasize that scattering processes analyzed in laboratories today are typically considered at zero temperature. Therefore, our results cannot be directly tested in current experiments. However, in astrophysical phenomena, temperature effects are significant and often dominant. Moreover, on a large scale, gravitational effects prevail, making gravitational processes like the one studied here both relevant and observable in such contexts.

\section*{Acknowledgments}

This work by A. F. S. is partially supported by National Council for Scientific and Technological Develo\-pment - CNPq project No. 312406/2023-1. L. A. S. E. thanks CAPES for financial support.

\section*{Data Availability Statement}

No Data associated in the manuscript.


\global\long\def\link#1#2{\href{http://eudml.org/#1}{#2}}
 \global\long\def\doi#1#2{\href{http://dx.doi.org/#1}{#2}}
 \global\long\def\arXiv#1#2{\href{http://arxiv.org/abs/#1}{arXiv:#1 [#2]}}
 \global\long\def\arXivOld#1{\href{http://arxiv.org/abs/#1}{arXiv:#1}}


\begin{thebibliography}{100}

\bibitem{Faraday} G. Cantor, ``Faraday's search for the gravelectric effect'',
\doi{} {Phys. Ed. {\bf 26}, 289 (1991).}

\bibitem{Maxwell} J. C. Maxwell, ``Philosophical Transact'',
\doi{} {Phil. Trans. Soc. Lond. {\bf 155}, 492 (1865).}

\bibitem{heaviside1893gravitational} O. Heaviside, ``A gravitational and electromagnetic analogy'',
\doi{} {The Electrician {\bf 31}, 281 (1893).}

\bibitem{lorentz1899considerations}H. A. Lorentz, ``Considerations on Gravitation''.  In: M. Janssen, J. D. Norton, J. Renn, T.  Sauer, J. Stachel (eds) ``The Genesis of General Relativity''. 
\doi{10.1007/978-1-4020-4000-9_13} {Boston Studies in the Philosophy of Science {\bf 250}. Springer, Dordrecht. (2007).} 

\bibitem{LT} J. Lense, H. Thirring, ``\"{U}ber den EinfluB der Eigenrotation der Zentralk\"{o}rper auf die Bewegung der Planeten und Monde nach der Einsteinschen Gravitationstheorie'',
\doi{}{Physikalische Zeitschrift {\bf 19}, 156 (1918).}

\bibitem{matte1953nouvelles} A. Matte, ``Sur De Nouvelles Solutions Oscillatoires Des Equations De La Gravitation'', 
\doi{10.4153/CJM-1953-001-3} {Canadian Journal of Mathematics {\bf 5},1 (1953).}

\bibitem{bel1958radiation} L. Bel, ``La radiation gravitationnelle''.
\doi{http://www.numdam.org/item/SJ_1958-1959__2__A12_0/} {Seminaire Janet. Mecanique analytique et mecanique celeste {\bf 2}, 26 (1958-1959).} 

\bibitem{Jair} J. Ramos, M. Montigny and F. C. Khanna, ``On a Lagrangian formulation of gravitoelectromagnetism'',
\doi{10.1007/s10714-010-0990-8} {Gen. Relativ. Gravit. {\bf 42}, 2403 (2010).}

\bibitem{jesus2022gravitational} W. D. R. Jesus, P. R. A.Souza, A. F. Santos and F. C. Khanna, ``Gravitational electron-positron scattering'', 
\doi{10.1140/epjp/s13360-022-02479-z} {Eur. Phys. J. Plus {\bf 137}, 260 (2022).}

\bibitem{santos2017gravitational} A. F. Santos and Faqir C. Khanna, ``Gravitational Bhabha scattering'',  
\doi{10.1088/1361-6382/aa89f4} {Class. Quantum Grav. {\bf 34}, 205007 (2017).}

\bibitem{alesandrogravitacional} A. F. Santos and F. C. Khanna, ``Gravitational M\"{o}ller scattering, Lorentz violation and finite temperature'',
\doi{10.1142/S0217732320502132} {Mod. Phys. Lett. A {\bf 35}, 26 (2020).}

\bibitem{Casimir} A. F. Santos and F. C. Khanna, ``Gravitational Casimir effect at finite temperature'',
\doi{ 10.1007/s10773-016-3156-y} {Int. J. Theor. Phys. {\bf 55}, 5356 (2016).}

\bibitem{Mashhon} B. Mashhoon, ``Gravitoelectromagnetism: A Brief Review'', 
\arXiv{gr-qc/0311030}{gr-qc}.

\bibitem{farrugia2020gravitoelectromagnetism} G. Farrugia, J. L. Said, and A. Finch, ``Gravitoelectromagnetism, solar system tests, and weak-field solutions in $f(T,B)$ gravity with observational constraints'', \doi{10.3390/universe6020034} {Universe {\bf 6}, 34 (2020).}

\bibitem{chatzistavrakidis2020torsion} A. Chatzistavrakidis, G. Karagiannis, and P. Schupp, ``Torsion-induced gravitational $\theta$ term and gravitoelectromagnetism'', \doi{10.1140/epjc/s10052-020-08600-9} {Eur. Phys. J. C {\bf 80}, 1085 (2020).}

\bibitem{bakopoulos2016gravitoelectromagnetism} A. Bakopoulos, ``Gravitoelectromagnetism: Basic principles, novel approaches and their application to Electromagnetism'', \doi{10.48550/arXiv.1610.08357} {arXiv:1610.08357 (2016).}


\bibitem{Compton} A. H. Compton, ``A Quantum Theory of the Scattering of X-Rays by Light Elements,''
\doi{10.1103/PhysRev.21.483} {Phys. Rev. {\bf 21}, 483 (1923).}

\bibitem{cabral} D. S. Cabral, A. F. Santos, ``Compton scattering in TFD formalism'', \doi{10.1140/epjc/s10052-023-11182-x}{{Eur. Phys. J. C} {\bf 83}, {25 (2023)}}.

\bibitem{Umezawa1}Y. Takahashi and H. Umezawa, ``Thermo Field Dynamics'', 
\doi{10.1142/S0217979296000817} {Int. Jour. Mod. Phys. B {\bf 10}, 1755 (1996).}

\bibitem{Umezawa2}Y. Takahashi, H. Umezawa and H. Matsumoto, {\it Thermofield Dynamics and Condensed States}, North-Holland, Amsterdan, (1982).

\bibitem{Book} F. C. Khanna, A. P. C. Malbouisson, J. M. C. Malboiusson and A. E. Santana, {\it Themal quantum field theory: Algebraic aspects and applications}, World Scientific, Singapore, (2009).

\bibitem{Umezawa22} H. Umezawa, {\it Advanced Field Theory: Micro, Macro and Thermal Physics}, AIP, New York, (1993).

\bibitem{Khanna1} A. E. Santana and F. C. Khanna, ``Lie groups and thermal field theory'',
\doi{10.1016/0375-9601(95)00394-I} {Phys. Lett. A {\bf 203}, 68 (1995).}

\bibitem{Khanna2} A. E. Santana, F. C. Khanna, H. Chu, and Y. C. Chang, ``Thermal Lie Groups, Classical Mechanics, and Thermofield Dynamics'',
\doi{10.1006/aphy.1996.0080} {Ann. Phys. {\bf 249}, 481 (1996).}

\bibitem{Santana1}A. E. Santana, A. Matos Neto, J. D. M. Vianna and F. C. Khanna, ``Symmetry groups, density-matrix equations and covariant Wigner functions'', 
\doi{10.1016/S0378-4371(99)00606-8} {Physica A: Statistical Mechanics and its Applications {\bf 280}, 405 (2000).}


\bibitem{Santana2}F. C. Khanna, A. P. C Malbouisson, J. M. C. Malbouisson and A. E. Santana, ``Thermoalgebras and path integral'',
\doi{10.1016/j.aop.2009.04.010} {Ann. Phys. {\bf 324}, 1931 (2009).}

\bibitem{matsubara} T. Matsubara, ``A new approach to quantum-statistical mechanics'', \doi{10.1143/PTP.14.351}{{Prog. Theor. Phys.} {\bf 14}, {351--378 (1955)}}.

\bibitem{schwinger1961brownian} J. Schwinger, ``Brownian motion of a quantum oscillator'', \doi{10.1063/1.1703727}{{J. Math. Phys.} {\bf 2}, {407--432 (1961)}}.

\bibitem{santos2017lorentz} A. F. Santos and F. C. Khanna, ``Lorentz violation in Bhabha scattering at finite temperature'', \doi{10.1103/PhysRevD.95.125012} {Phys. Rev. D {\bf 95}, 125012 (2017).}

\bibitem{santos2016quantized} A. F. Santos and F. C. Khanna, ``Quantized gravitoelectromagnetism theory at finite temperature'', \doi{10.1142/S0217751X16501220} {Int. J. Mod. Phys. A {\bf 31}, 1650122 (2016).}

\bibitem{cabral2023thermal} D. S. Cabral, A. F. Santos, and R. Bufalo, ``Thermal pair production from photon-photon collision: Breit--Wheeler process at finite temperature'', \doi{10.1140/epjc/s10052-023-12281-5} {The European Physical Journal C {\bf 83}, 1113 (2023).}

\bibitem{cabral2024lorentz} D. S. Cabral, L. A. S. Evangelista, J. C. R. de Souza, L. H. A. R. Ferreira, and A. F. Santos, ``Lorentz-violating Yukawa theory at finite temperature'', \doi{10.1103/PhysRevD.110.095022} {Physical Review D {\bf 110}, 095022 (2024).}

\bibitem{Filipe} L. Filipe Costa and Carlos A. R. Herdeiro, ``Gravitoelectromagnetic analogy based on tidal tensors'',
\doi{10.1103/PhysRevD.78.024021} {Phys. Rev. D {\bf 78}, 024021 (2008). }

\bibitem{Maartens}R. Maartens and B. A. Bassett, ``Gravito-electromagnetism'', 
\doi{10.1088/0264-9381/15/3/018} {Class. Quant. Grav. {\bf 15}, 705 (1998).}

\bibitem{danehkar2009significance} A. Danehkar, ``On the Significance of the Weyl Curvature in a Relativistic Cosmological Model'',
\doi{10.1142/S0217732309032046} {Mod. Phys. Lett. A {\bf 24}, 3113 (2009).}

\bibitem{misner1973k} C. W. Misner, ``K, S. Thorne, J. A. Wheeler, Gravitation, vol. 1'', \doi{} {W. H. Freeman Company, San Francisco (1973).}


\bibitem{das2023finite} A. Das, ``Finite temperature field theory'', \doi{} {World Scientific (2023).}

\bibitem{xu1995compton} H.-H. Xu and C.-H. Xu, ``Compton scattering at finite temperature'', \doi{10.1103/PhysRevD.52.6116} {Phys. Rev. D {\bf 52}, 6116 (1995).}





\end{thebibliography}
\end{document}